\documentclass[english,prb, reprint, twocolumn,showpacs]{revtex4}
\usepackage[T1]{fontenc}
\usepackage[latin9]{inputenc}
\usepackage{array}
\usepackage{multirow}
\usepackage{tipa}
\usepackage{tipx}
\usepackage{amsmath}
\usepackage{graphicx}

\makeatletter

\providecommand{\tabularnewline}{\\}

\@ifundefined{textcolor}{}
{%
 \definecolor{BLACK}{gray}{0}
 \definecolor{WHITE}{gray}{1}
 \definecolor{RED}{rgb}{1,0,0}
 \definecolor{GREEN}{rgb}{0,1,0}
 \definecolor{BLUE}{rgb}{0,0,1}
 \definecolor{CYAN}{cmyk}{1,0,0,0}
 \definecolor{MAGENTA}{cmyk}{0,1,0,0}
 \definecolor{YELLOW}{cmyk}{0,0,1,0}
}

\makeatother

\usepackage{babel}
\begin{document}

\title{Selection Rules for Non-Radiative Carrier Relaxation Processes in Semiconductor Quantum Dots}

\author{E. R. Schmidgall}
\thanks{Current address: Department of Physics, University of Washington, Seattle, Washington 98195, USA}
\email{eschmid@uw.edu}
\affiliation{Department of Physics and the Solid State Institute, The Technion
- Israel Institute of Technology, Haifa 32000, Israel}

\author{Y. Benny}

\affiliation{Department of Physics and the Solid State Institute, The Technion
- Israel Institute of Technology, Haifa 32000, Israel}

\author{I. Schwartz}

\affiliation{Department of Physics and the Solid State Institute, The Technion
- Israel Institute of Technology, Haifa 32000, Israel}

\author{R. Presman}

\affiliation{Department of Physics and the Solid State Institute, The Technion - Israel Institute of Technology, Haifa 32000, Israel}

\author{L. Gantz}

\affiliation{Department of Physics and the Solid State Institute, The Technion
- Israel Institute of Technology, Haifa 32000, Israel}

\author{Y. Don}

\affiliation{Department of Physics and the Solid State Institute, The Technion - Israel Institute of Technology, Haifa 32000, Israel}

\author{D. Gershoni}

\affiliation{Department of Physics and the Solid State Institute, The Technion
- Israel Institute of Technology, Haifa 32000, Israel}

\begin{abstract}
Time resolved intensity cross-correlation measurements of radiative cascades 
are used for studying non-radiative relaxation processes of excited carriers confined in semiconductor 
quantum dots. We spectrally identify indirect radiative cascades which include intermediate 
phonon assisted relaxations. The energy of the first photon reveals the multicarrier 
configuration prior to the non-radiative relaxation, while the energy of the 
second photon reveals the configuration after the relaxation. The intensity cross correlation measurements thus
provide quantitative measures of the non-radiative processes and their selection rules. 
We construct a model which accurately describes the experimental observations
in terms of the electron-phonon and electron-hole exchange interactions. 
Our measurements and model provide a new tool for engineering 
relaxation processes in semiconductor nanostructures. 
\end{abstract}


\maketitle
\section{Introduction}
Semiconductor quantum dots (QDs) have received considerable attention
due to their atomic-like spectral features and their compatibility
with modern microelectronics technology. These QDs create a three dimensional potential
well confining charge carriers in all spatial directions.
The confinement results in a discrete spectrum of energy levels and
energetically sharp optical transitions between these levels. As such,
QDs are considered promising building blocks for future technologies
involving single and correlated photon emitters, single photon detectors and various other platforms for quantum information processing
(QIP).\cite{imamoglu1999,loss1998,akopian2006,santori2001,muller2014} 

QDs are particularly useful for these applications since they provide excellent interfaces between photons (flying qubits)\cite{turchette1995,knill2001}
and confined charge carrier spins (anchored matter qubits).\cite{kosaka2008,benny2011prl,degreve2011}
QD-confined charge carrier spins have been used as implementations
of qubits, as quantum gates,\cite{benny2011prl,kosaka2009,press2010,kodriano2012}
and entanglement between a QD spin and a photon has been demonstrated.\cite{akopian2006,degreve2012,gao2012,shaibley2013}

Unlike isolated atoms and molecules, however, QDs are strongly coupled to their environment. This coupling on 
one hand facilitates their applications in contemporary technology, but on the other hand it leads to unavoidable non-radiative 
relaxation and decoherence processes, which involve lattice vibrations (phonons).  Indeed, while quasiresonant electron-LO phonon interaction in single QDs has been reported and modeled previously \cite{hameau1999,stauber2000,melinkov2001,kaczmarkiewicz2010}, further development of QDs for various applications in QIP continues to require understanding and control of phonon-assisted processes. 

\begin{figure}
\includegraphics[width=1\columnwidth]{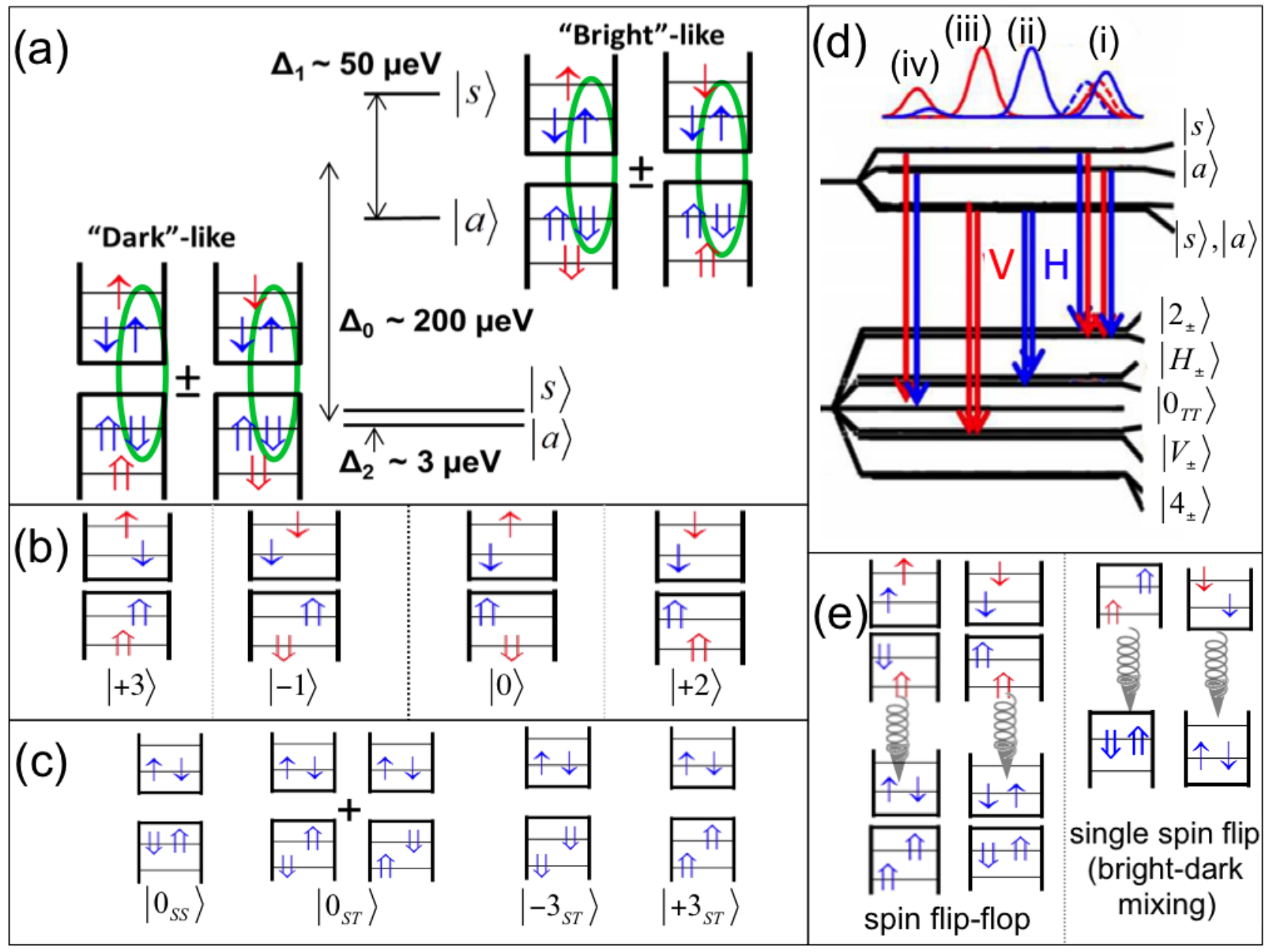}
\protect\caption{(a) Schematic illustration of the QD-confined ground-state triexciton.
$\uparrow(\Uparrow)$ represents an electron (heavy hole). Spin projection
corresponds to arrow direction. Green ovals indicate recombining electron-hole pairs that result in the excited biexciton states shown in (b). (b) Schematic illustration of resulting excited biexciton states after recombination of a ground e-h pair  (indicated by a green oval in (a)) from the ``bright''-like (``dark''-like) $XXX^0$. Only one (out of two, equivalent) spin projection is shown for clarity. The notation corresponds to that in Equation \ref{eq:basis_states}. (c) Lower-energy biexcitonic eigenstates that give rise to distinct PL emission lines as seen in Fig. \ref{fig:PL}. The notation here corresponds to that used in Equation \ref{eq:ground_states}. (d) The e-triplet--h-triplet biexcitonic eigenstates ordered by their enenergies and the optical transitions to these eigenstates from the triexciton eigenstates. (e) Schematic descriptions of the two non-radiative processes considered here--the spin flip-flop process and a spin flip due to dark-bright exciton mixing--showing their effect on two different excited biexciton states.
\label{fig:theory}}
\end{figure}

Here, we experimentally identify and theoretically explain relaxation
mechanisms for QD-confined multicarrier spin configurations containing two
electrons and two heavy holes. For studying these mechanisms,
we use temporally-resolved, polarization-sensitive intensity cross-correlation
measurements of two-photon radiative cascades initiating from a QD-confined 
triexciton--three electron-hole (e-h) pairs (Fig. \ref{fig:theory}a).
Recombination of one ground level e-h pair out of the three pairs of the triexciton leaves in the QD two
e-h pairs, a biexciton, in an excited state  (Fig. \ref{fig:theory}b).
The relaxation of the excited biexciton can then be studied experimentally by
detecting emission due to a second e-h pair recombination, 
which occurs after the excited biexciton relaxes non-radiatively to various lower energy biexciton configurations (Fig. \ref{fig:theory}d).

The most studied biexciton configuration is its ground level, in which both 
the two electrons and the two heavy holes form 
spin singlets in their respective ground levels. Since both types of carriers
are in their ground levels, further relaxation is only possible via 
radiative recombination.~\cite{regelman2001,akopian2006} 
In our QDs, we typically observe three additional spectral lines which result from 
biexciton recombinations.
These lines result from metastable biexciton configurations in which the two 
electrons form a spin singlet in their ground level 
but the two holes, one in the ground level and one in an excited level, 
form spin triplets (Fig. \ref{fig:theory}d). 
These configurations are spin blockaded 
from further phonon assisted relaxations and therefore result in distinct photoluminescence (PL) emission lines (Fig. \ref{fig:PL}).~\cite{kodriano2010}

We note here that there is an inherent asymmetry between electrons (which relax to their ground level faster than the radiative rate) and holes (which do not). This asymmetry was explained by Benny {\it et al.} \cite{benny2014} in terms of longitudinal optical (LO) phonon-induced mixing between the first two electron levels ($1e$ and $2e$). These levels are efficiently mixed by the 
e-LO phonon interaction \cite{kash1985,kash1987}, since the energy difference between these levels ($\sim30$ meV) closely resonates with the energy of an LO phonon in these compounds ($\sim 32$ meV)\cite{heitz1999,findeis2000,lemaitre2001,sarkar2005}.  In contrast, the energy difference between the first two hole levels ($1h$,$2h$ $\sim 10$ meV) is considerably smaller than the LO phonon energy, and thereby the mixing between these levels is negligibly small. 

The radiative cascades that we study are such that the first detected photon with well-defined energy and polarization fully characterizes the initial excited biexciton configuration, while the second detected photon fully characterizes the final biexciton configuration to which the non-radiative relaxation occurs. Thus, polarization-sensitive intensity cross-correlation measurements provide an excellent probe for the
various phonon assisted relaxation processes.

In order to understand the experimental observations, 
we developed a theoretical model in which the Fr\"ohlich electron - LO phonon interaction 
and the e-h exchange interaction~\cite{benny2014} including terms which result from the QD deviation from symmetry~\cite{zielinski2015} 
and lead to dark exciton - bright exciton mixing, are added as perturbations to the QD multi-carrier Hamiltonian.  
The insight thus gained suggests novel ways for 
engineering semiconductor QDs structurally, or for tuning their properties by the application
of external stress, electric, or magnetic fields. In this manner, control of the resulting non-radiative
relaxation processes may thereby achieve deterministic spin relaxation channels.   

Section \ref{sec:sample} presents our sample and experimental setup. Section \ref{sec:states} discusses the triexcitonic and biexciton states included in the model and the radiative and non-radiative transitions between these states. Section \ref{sec:model} presents the Hamiltonian for the electron-LO phonon and e-h exchange interactions and includes the impact of reduced QD symmetry. Finally, Section \ref{sec:exp} presents the experimentally measured second-order intensity correlation functions that agree quantitatively with the predictions of the model.

\section{Sample and Experimental Setup}
\label{sec:sample}

The sample that we study was grown by molecular-beam epitaxy on a
(001)-oriented GaAs substrate. One layer of strain-induced InGaAs
QDs was grown in a planar microcavity formed by two distributed Bragg
reflecting mirrors, and the microcavity was optimized for the range
of wavelengths corresponding to PL emission caused
by optical recombination between ground-state carriers in these QDs.\cite{poem2007,kodriano2010,benny2012,benny2011prb}
The measurements were carried out in a $\mu$-PL setup at 4.2 K. The
setup provides spatial resolution of about 1 $\mu$m, spectral resolution
of about 10 $\mu$eV and temporal resolution of about 400 ps in measuring
the arrival times of up to four photons originating from four spectral
lines at given polarizations. More details about the sample \cite{garcia1997,dekel2001}
and the experimental setup \cite{poemprb2010,benny2011prb} are given
in earlier publications. 

\section{States and Transitions} 
\label{sec:states}

Our discussion follows the notations of Benny {\it et al.}~\cite{benny2011prb}: a single carrier
state is described by its envelope wavefunction or orbital mode
$\left(O=1,2\right)$ where the number represents the energy
order of the level such that $O=1$ represents the ground state. $O$
is followed by the type of carrier ($e$, electron; $h$, heavy hole)
and a superscript describing the occupation of the single-carrier
state. The total electron and hole state spin projection on the QD symmetry axis is  described by corresponding subscripts.   

In Fig. \ref{fig:theory}(a), we present the four possible spin configurations of the ground state triexciton. \cite{schmidgall2014}
Its unpaired, higher-energy electron and hole can be either
in a mutual spin-parallel ("dark like") or a spin anti-parallel ("bright like" ) configuration (Fig. \ref{fig:theory}a).
Consequently, the fine structure of the triexciton is similar
to that of the exciton.\cite{bayer2002,takagahara2000,poem2007,benny2011prl, schmidgall2014}
 In our notation, the 
 "bright" ["dark"-] like ground triexciton states are given by: $(1e^{2}2e^{1})_{\pm\frac{1}{2}}(1h^{2}2h^{1})_{\mp \frac{3}{2}}$ [$(1e^{2}2e^{1})_{\pm \frac{1}{2}}(1h^{2}2h^{1})_{\pm\frac{3}{2}}$].

We limit our studies to triexciton radiative recombinations of ground level e-h pairs, 
since only these recombinations result in excited biexciton configurations.
Altogether, there are in total $2^{4}=16$ possible excited biexciton configurations~\cite{dekel2000,benny2011prb}.
In seven of these configurations, either the electrons, or the holes, or both, form a singlet spin state.
In these cases, the relaxation to the singlet ground level is spin preserving and happens quite fast, on a picosecond time scale.~\cite{schwartz2015b, poemprb2010, benny2011prb}
As a result, the spectral width of the optical transitions to these states are rather broad, rendering the spectral identification 
of the optical transition quite challenging. 
In contrast, the remaining nine configurations in which both the electrons and holes form spin triplet states
are spin blockaded  for further relaxation to the ground biexciton level. Consequently, the optical transitions to these levels are 
spectrally narrow and can be easily identified.~\cite{schmidgall2014}

In our notation these excited biexcitonic states 
are $\left(1e^{1}2e^{1}\right)_{T^e}\left(1h^{1}2h^{1}\right)_{T^h}$
where $T^{e}(T^{h})$ represents one of the three electron (hole) spin triplet states, $T^{e}_{-1,0,1}$ ($T^{h}_{-3,0,3}$).
We denote each basis state by its total angular momentum projection on the QD growth direction ($\hat{z}$). 

\begin{align}
\left|\pm 4\right\rangle =&\ T^{e}_{\pm 1}T^{h}_{\pm 3}\nonumber \\ 
\left|\pm 3\right\rangle =&\ T^{e}_{0}T^{h}_{\pm 3}\nonumber \\
\left|\pm 2\right\rangle =&\ T^{e}_{\mp 1}T^{h}_{\pm 3}\label{eq:basis_states}\\ 
\left|\pm 1\right\rangle =&\ T^{e}_{\pm 1}T^{h}_{0}\nonumber \\
\left|0\right\rangle =&\ T^{e}_{0}T^{h}_{0} \nonumber
\end{align}
For calculating the excited biexciton eigenstates and the selection rules for optical transitions to these states we used a 
many-carrier Hamiltonian which includes Coulomb and exchange (isotropic and anisotropic) 
interactions between the electrons and the holes.\cite{poem2007, benny2011prb} The diagonalized states of this many-carrier Hamiltonian are shown in Fig. \ref{fig:theory}(c). We will use the notation 
\begin{align}
&\left| 2_{\pm} \right \rangle =1/\sqrt{2}\left[ \left|2 \right\rangle \pm \left|-2\right\rangle \nonumber\right] \\
&\left|H_{\pm }\right \rangle =1/2\left[ \left(\left|1 \right\rangle \pm \left|-1 \right \rangle \right) + \left(\left|3 \right\rangle \pm \left|-3 \right \rangle \right)\right] \nonumber \\
&\left| 0_{TT} \right \rangle = \left| 0 \right\rangle \label{eq:diag_states} \\
&\left|V_{\pm}\right \rangle =1/2\left[\left(\left|1 \right\rangle \pm \left|-1 \right \rangle \right) - \left(\left|3 \right\rangle \pm \left|-3 \right \rangle \right)\right] \nonumber \\
&\left| 4_{\pm}\right \rangle = 1/\sqrt{2}\left[\left|4 \right\rangle \pm \left|-4\right\rangle \right]\nonumber 
\end{align}

We note here that of these 9 states, only 7 are optically accessible. The two $|4_{\pm}\rangle$ states are completely dark.

In Fig. \ref{fig:theory}(d), we display the 10 possible optical transitions between the 4 triexciton states and these 9 e-triplet--h-triplet excited biexciton states.

\begin{figure}[b]
\begin{center}
\includegraphics[width=\columnwidth]{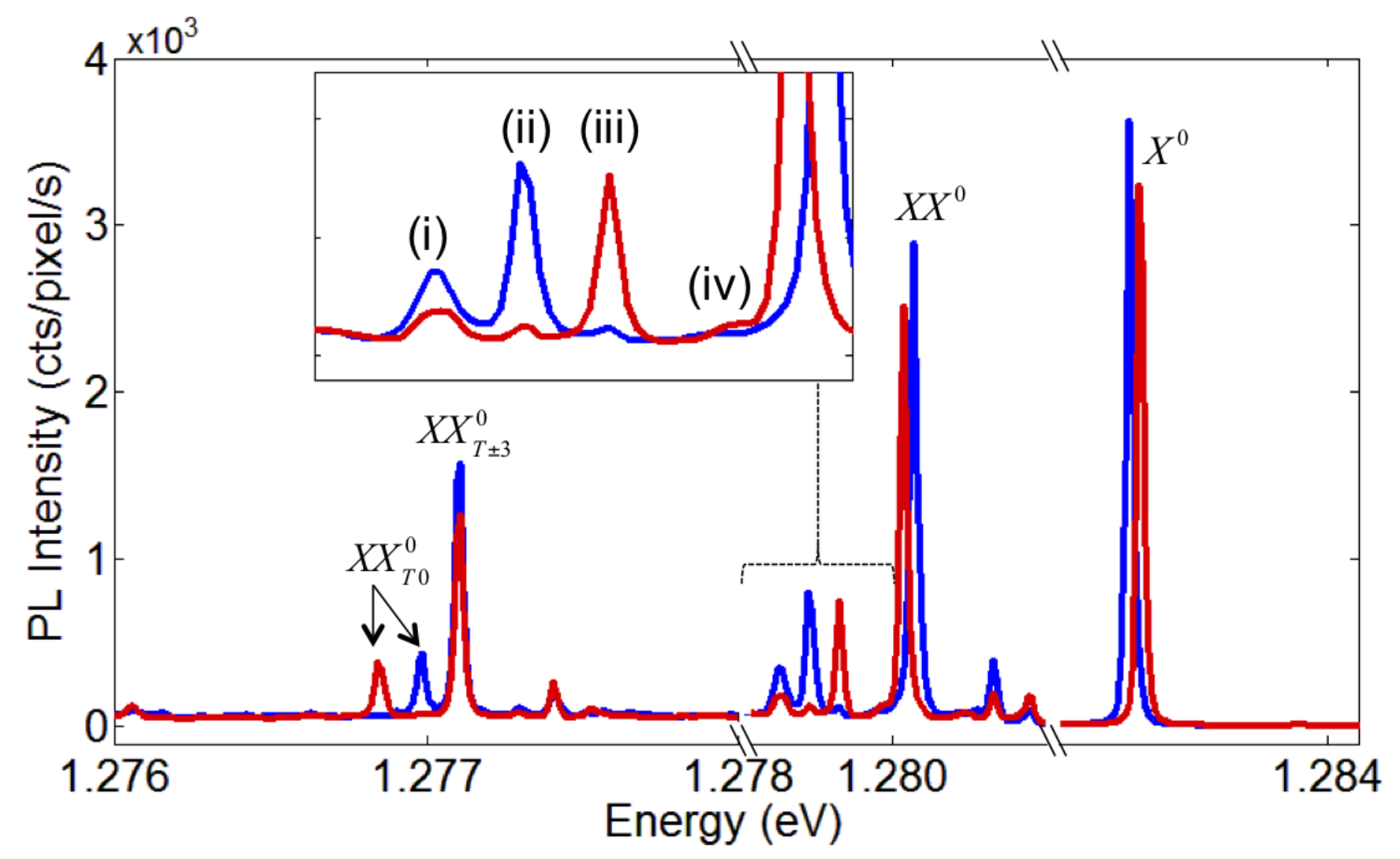}
\caption{PL spectrum of a single QD under nonresonant excitation with a 455nm cw laser. The emission lines are indicated using the notation from Eq. \ref{eq:ground_states} or by a Roman numeral matched to the transitions depicted in Fig. \ref{fig:theory}(d).}
\label{fig:PL}
\end{center}
\end{figure}

The four spectrally identified biexciton configurations in order of decreasing energy are:
\begin{align}
&\left|0_{ST}\right\rangle =1e^2T^{h}_0\nonumber \\  
&\left|\pm3_{ST}\right\rangle =1e^2[T^{h}_{\pm3}]\label{eq:ground_states}\\
&\left|0_{SS}\right\rangle=1e^21h^2\nonumber
\end{align}
The lowest energy state is the ground biexciton state 
in which both the two electrons and the two holes form spin singlet states in their respective ground level.
Recombination of a ground e-h pair from this state gives rise to the two cross-rectilinearly polarized $XX^0$ spectral lines.
The additional three biexciton states which give rise to emission lines are states in which the two electrons form 
 ground level singlet states, but the two holes form metastable triplet 
states. These biexciton states $|\pm 3_{ST}\rangle$, and $|0_{ST}\rangle$ give rise to the unpolarized 
$XX^0_{T_{\pm3}}$ line, and the two cross rectilinearly polarized $XX^0_{T_0}$ lines, respectively \cite{kodriano2010}. These emission lines are identified in Fig. \ref{fig:PL}. 

The spin wavefunctions and energy levels of these ground biexciton configurations are schematically described in Fig. \ref{fig:theory}(c).
From the above discussion, and the fact that phonons cannot alter the relaxing electron spin state it follows that these ground biexciton levels are phonon coupled to the following excited levels, in which the two
electrons form an excited singlet spin state:  

\begin{align}
&\left|0^{*}_{ST}\right\rangle =(1e^12e^1)_ST^{h}_0 \nonumber \\ 
&\left|\pm 3^{*}_{ST}\right\rangle =(1e^12e^1)_S [T^{h}_{\pm3}]\label{eq:excited_states}\\
&\left|0^{*}_{SS}\right\rangle=(1e^12e^1)_S1h^2\nonumber
\end{align}

The non-radiative relaxation processes from the excited biexciton configurations in Fig. \ref{fig:theory}(b)
to the ground biexciton configurations in Fig. \ref{fig:theory}(c) are schematically represented in Fig. \ref{fig:theory}(e) by downward curly arrows.
These non-radiative electronic transitions are mediated by phonons,\cite{poemprb2010, kodriano2010, benny2014} which, for the most part, preserve the electronic spin.\cite{kodriano2010, poemprb2010}
Clearly, however, the e-triplet-h-triplet excited biexciton 
configurations relax to the ground biexciton configurations without preserving the spin.
The quantitative study of these spin non-conserving phonon mediated relaxations is the subject that we study in detail below.   

\begin{table}[bt]
\begin{tabular}{|c c|}
\hline 
{\bf Basis State(s)} & {\bf Energy} \\\hline \hline
$T^{e}T^{h}, 0LO$ (Eqn. \ref{eq:basis_states}) & $E_{carr}^{*}-E_{1e2e1e2e}^{exch}-E_{1h2h1h2h}^{exch}$\footnote{$E_{carr}^{*}=E_{2h}+E_{2e}+E_{g}+E^{Coul}_{2e1e1e2e}+E^{Coul}_{2h1h1h2h}-E^{Coul}_{1e1h1h1e}-E^{Coul}_{2e1h1h2e}-E^{Coul}_{1e2h2h1e}-E^{Coul}_{2e2h2h2e}.$}\\
$|\pm 3_{ST}^{*},0LO\rangle$ (Eqn. \ref{eq:excited_states}) &$E_{carr}^{*}+E_{1e2e1e2e}^{exch}-E_{1h2h1h2h}^{exch}$\\
$|0^{*}_{ST},0LO\rangle$ (Eqn. \ref{eq:excited_states}) &$E_{carr}^{*}+E_{1e2e1e2e}^{exch}-E_{1h2h1h2h}^{exch}$\\
$|0^{*}_{SS},0LO\rangle$ (Eqn. \ref{eq:excited_states})&$E_{carr}^{*S}+E_{1e2e1e2e}^{exch}$\footnote{$E_{carr}^{*S}=E_{1h}+E_{2e}+E_{g}+E^{Coul}_{2e1e1e2e}+E^{Coul}_{1h1h1h1h}-2E^{Coul}_{1e1h1h1e}-2E^{Coul}_{2e1h1h2e}$}\\
$|\pm 3_{ST},1LO\rangle$ (Eqn. \ref{eq:ground_states}) &$E_{carr}-E_{1h2h1h2h}^{exch}+E_{LO}$\footnote{$E_{carr}=E_{2h}+E_{1e}+E_{g}+E^{Coul}_{1e1e1e1e}+E^{Coul}_{1h2h2h1h}-2E^{Coul}_{1e1h1h1e}-2E^{Coul}_{1e2h2h1e}$}\\
$|0_{ST},1LO\rangle $ (Eqn. \ref{eq:ground_states}) &$E_{carr}-E_{1h2h1h2h}^{exch}+E_{LO}$\\
$|0_{SS},1LO\rangle $ (Eqn. \ref{eq:ground_states}) &$E_{carr}^{S}+E_{LO}$\footnote{$E_{carr}^{S} = E_{1h}+E_{1e}+E_{g}+E^{Coul}_{1e1e1e1e}+E^{Coul}_{1h1h1h1h}-4E^{Coul}_{1e1h1h1e}$}\\ 
\hline
\end{tabular}
\caption{The basis states and energies of the $H_{0}$ Hamiltonian}
\label{tbl:states} 
\end{table}

\section{The Theoretical Model for Phonon Mediated Relaxations} 
\label{sec:model}

To study the spin non-conserving phonon mediated relaxations, we constructed an effective Hamiltonian
\begin{equation}
H=H_{0}+H_{e-h}+H_{e-LO}+H_{DB}
\end{equation}
where $H_{0}$ contains the single-carrier part of the electrons and holes,
the electron-electron and hole-hole exchange interactions, the electron-hole
direct Coulomb interactions, and the single phonon Hamiltonian.
The diagonalized basis for this Hamiltonian are the 7 optically active functions of the e-triplet-h-triple states (Eqn. \ref{eq:basis_states}),  
 the 4 e-singlet ground biexciton configurations (Eqn. \ref{eq:ground_states}), with 1 LO phonon, each, and their 4 excited states (Eqn. \ref{eq:excited_states}), with no LO phonon. These states, including those states containing an LO phonon, are treated as discrete states. Table \ref{tbl:states} shows these basis states and their energies ($H_{0}$). 
$H_{e-h}$ contains the electron-hole exchange interaction obtained in this basis.

For a $C_{2v}$ symmetric QD, the anisotropic e-h exchange interaction does not mix between dark and bright excitons. In this case the terms in $H_{e-h}$ connect between states which differ by an electron and a hole spin direction.
It therefore mixes between states which differ by two units of angular momentum.  
This interaction therefore leads to relaxation terms resulting in an e-h spin flip-flop.~\cite{benny2014}

Benny {\it et al.}~\cite{benny2011prb} expressed $H_{e-h}$ and obtained the eigenenergies and eigenfunctions of the e-triplet--h-triplet biexciton states  
depicted in Eqs. \ref{eq:basis_states} and \ref{eq:diag_states} and in Fig. \ref{fig:theory}(d). 
Since the e-h exchange interaction is more than an order of magnitude
smaller than the e-e and the h-h exchange interactions, the spin flip-flop relaxation rates are typically much slower than the radiative rate.~\cite{benny2011prb} 
 
When the QD deviates from $C_{2v}$ symmetry, there are mixing terms between the dark and bright
exciton subspaces.  These terms can be viewed as relaxation terms in which either a single electron or a single hole flips its spin (spin-flip interactions). This situation was discussed recently by Zielinski {\it et al.}~\cite{zielinski2015}, who showed that the e-h exchange Hamitonian for exciton between the $i^{th}$ electronic level to the $j^{th}$ hole level is given by
{\small\begin{equation}
H_{X^0_{i,j}}=\frac{1}{2}
\left(\begin{array}{r|cccc}
&1&-1&2&-2\\
\hline
1&\Delta_{0}^{i,j}&\Delta_{1}^{i,j}&\Delta_{e}^{i,j}&\Delta_{h}^{i,j}\\
-1&\Delta_{1}^{i,j}&\Delta_{0}^{i,j}&\Delta_{h}^{i,j}&\Delta_{e}^{i,j}\\
2&\Delta_{e}^{i,j}&\Delta_{h}^{i,j}&-\Delta_{0}^{i,j}&\Delta_{2}^{i,j}\\
-2&\Delta_{h}^{i,j}&\Delta_{e}^{i,j}&\Delta_{2}^{i,j}&-\Delta_{0}^{i,j}\\
\end{array}\right)
\end{equation}}
where $\Delta_{0,1,2}^{i,j}$, are the usual e-h exchange interaction terms for a $C_{2v}$ symmetrical QD
and $\Delta_{e}^{i,j}$ ($\Delta_{h}^{i,j}$) is a mixing term which connect between states of opposite electron (hole) exciton states. 
Typical magnitudes for these mixing terms were calculated in Ref~\cite{zielinski2015}, and found to be only a small fraction of the e-h anisotropic exchange interaction $\Delta_1^{i,j}$ . 
Therefore, single carrier spin flip relaxation rates are expected to be even slower than spin flip-flop rates. Here, these single carrier spin flip terms are incorporated into $H_{DB}$. 

$H_{e-LO}$ contains the effect of the electron LO phonon interaction. Since phonons do not 
 interact with the electronic spin, $H_{e-LO}$ only mixes between similar spin configurations, which belong to different electron orbitals. 
Typically, therefore, $H_{e-LO}$ leads to spin preserving single carrier relaxation terms.
 However, in Ref. \cite{benny2014} it was shown that when the LO-phonon energy is comparable 
to the single carrier energy level separation,
the combined effect of $H_{e-LO}$ and $H_{e-h}$ lead to pronounced spin flip-flop relaxation rates, which become comparable to, and even faster than, the radiative rate.  We show below that this is true also for single carrier spin flip processes, which result from the bright-dark exciton mixing induced by the symmetry reduction. 

The $H_{eh,LO,DB}=H_{e-h}+H_{e-LO}+H_{DB}$ Hamiltonian in the basis $|+2\rangle$, $|-1\rangle$, $|+3\rangle$, $|0\rangle$, $|-3\rangle$, $|+1\rangle$, $|-2\rangle$, $|+3_{ST}^{*}\rangle$, $|0^{*}_{ST}\rangle$, $|-3_{ST}^{*}\rangle$, $|0^{*}_{SS}\rangle$, $|+3_{ST},1LO\rangle$, $|0_{ST},1LO\rangle$, $|-3_{ST},1LO\rangle$, $|0_{SS},1LO\rangle$ is therfore given by: 

\begin{widetext}
{\tiny
\begin{align}H_{eh,LO,DB}&=\nonumber\\
&\frac{1}{2}\left( \begin{array}{ccccccccccccccc}\tilde{\Delta}_{0+}&4\tilde{\Delta}_{h}&4\tilde{\Delta}_{e}&\tilde{\Delta}_{1+}&0&0&0&0&-\tilde{\Delta}_{1-}&0&\tilde{\Delta}_{1SS}&-2\tilde{\Delta}_{e}&0&0&0\\
4\tilde{\Delta}_{h}&0&\tilde{\Delta}_{2+}&8\tilde{\Delta}_{e}&\tilde{\Delta}_{1+}&0&0&\tilde{\Delta}_{2-}&0&-\tilde{\Delta}_{1-}&0&0&-4\tilde{\Delta}_{e}&0&0\\ 
4\tilde{\Delta}_{e}&\tilde{\Delta}_{2+}&0&8\tilde{\Delta}_{h}&0&\tilde{\Delta}_{1+}&0&-\tilde{\Delta}_{0-}&0&0&0&0&0&0&0\\
\tilde{\Delta}_{1+}&8\tilde{\Delta}_{e}&8\tilde{\Delta}_{h}&0&8\tilde{\Delta}_{h}&8\tilde{\Delta}_{e}&\tilde{\Delta}_{1+}&0&0&0&\tilde{\Delta}_{0SS}&0&0&0&0\\
0&\tilde{\Delta}_{1+}&0&8\tilde{\Delta}_{h}&0&\tilde{\Delta}_{2+}&4\tilde{\Delta}_{e}&0&0&\tilde{\Delta}_{0-}&0&0&0&0&0\\
0&0&\tilde{\Delta}_{1+}&8\tilde{\Delta}_e&\tilde{\Delta}_{2+}&0&4\tilde{\Delta}_h&\tilde{\Delta}_{1-}&0&-\tilde{\Delta}_{2-}&0&0&4\tilde{\Delta}_e&0&0\\
0&0&0&\tilde{\Delta}_{1+}&4\tilde{\Delta}_e&4\tilde{\Delta}_h&\tilde{\Delta}_{0+}&0&\tilde{\Delta}_{1-}&0&\tilde{\Delta}_{1SS}&0&0&2\tilde{\Delta}_e&0\\
0&\tilde{\Delta}_{2-}&-\tilde{\Delta}_{0-}&0&0&\tilde{\Delta}_{1-}&0&0&8\tilde{\Delta}_{h}&0&-4\tilde{\Delta}_{h}&C_{F}&-4\tilde{\Delta}_{h}&0&2\tilde{\Delta}_{h}\\
-\tilde{\Delta}_{1-}&0&0&0&0&0&\tilde{\Delta}_{1-}&8\tilde{\Delta}_{h}&0&8\tilde{\Delta}_{h}&0&-4\tilde{\Delta}_{h}&C_F&-4\tilde{\Delta}_{h}&0\\
0&-\tilde{\Delta}_{1-}&0&0&\tilde{\Delta}_{0-}&-\tilde{\Delta}_{2-}&0&0&8\tilde{\Delta}_{h}&0&4\tilde{\Delta}_{h}&0&-4\tilde{\Delta}_{h}&C_F&-2\tilde{\Delta}_{h}\\
\tilde{\Delta}_{1SS}&0&0&\tilde{\Delta}_{0SS}&0&0&\tilde{\Delta}_{1SS}&-4\tilde{\Delta}_{h}&0&4\tilde{\Delta}_{h}&0&2\tilde{\Delta}_{h}&0&-2\tilde{\Delta}_{h}&C_F\\
-2\tilde{\Delta}_e&0&0&0&0&0&0&C_F&-4\tilde{\Delta}_{h}&0&2\tilde{\Delta}_{h}&0&4\tilde{\Delta}_{h}&0&-2\tilde{\Delta}_{h}\\
0&-4\tilde{\Delta}_e&0&0&0&4\tilde{\Delta}_e&0&-4\tilde{\Delta}_{h}&C_F&-4\tilde{\Delta}_{h}&0&4\tilde{\Delta}_{h}&0&4\tilde{\Delta}_{h}&0\\
0&0&0&0&0&0&2\tilde{\Delta}_e&0&-4\tilde{\Delta}_{h}&C_F&-2\tilde{\Delta}_{h}&0&4\tilde{\Delta}_{h}&0&2\tilde{\Delta}_{h}\\
0&0&0&0&0&0&0&2\tilde{\Delta}_{h}&0&-2\tilde{\Delta}_{h}&C_F&-2\tilde{\Delta}_{h}&0&2\tilde{\Delta}_{h}&0\\
\end{array} \right)\label{eq:hamfull}\end{align}}
\end{widetext}

where {\small{}$$\tilde{\Delta}_{0\pm}=\left(\Delta_{0}^{1e,1h}+\Delta_{0}^{1e,2h}\pm \Delta_{0}^{2e,1h}\pm\Delta_{0}^{2e,2h}\right)/4$$}
and likewise, {\small{}$$\tilde{\Delta}_{1\pm,2\pm}=\left(\Delta_{1,2}^{1e,1h}+\Delta_{1,2}^{1e,2h}\pm \Delta_{1,2}^{2e,1h}\pm \Delta_{1,2}^{2e,2h}\right)/8.$$} For the interaction with the singlet states, {\footnotesize{}\begin{displaymath}\tilde{\Delta}_{(0,1)SS} = \left(\Delta_{(0,1)}^{1e,1h}-\Delta_{(0,1)}^{1e,2h}-\Delta_{(0,1)}^{2e,1h}+\Delta_{(0,1)}^{2e,2h}\right)/(4,8).\end{displaymath}}
Likewise,
{\small{}$$\tilde{\Delta}_{(e,h)}=\left(\Delta_{(e,h)}^{1e,1h}+\Delta_{(e,h)}^{1e,2h}+ \Delta_{(e,h)}^{2e,1h}+\Delta_{(e,h)}^{2e,2h}\right)/4$$}.

$C_{F}$ represents the Fr{\"o}hlich coupling between the
excited electron and the LO phonon. 
A Fr{\" o}hlich coupling constant 
of 6.4 meV 
was used.~\cite{benny2014} 
The values of $\Delta_{0},$
$\Delta_{1}$, and $\Delta_{2}$ for various configurations of carriers
have been deduced in previous works.~\cite{poem2007,benny2014,benny2011prb,benny2012}  

Table \ref{tab:parameters} lists and references the values of all the parameters used in our model calculations. 

\begin{figure}[tb]
\begin{center}
\includegraphics[width=\columnwidth]{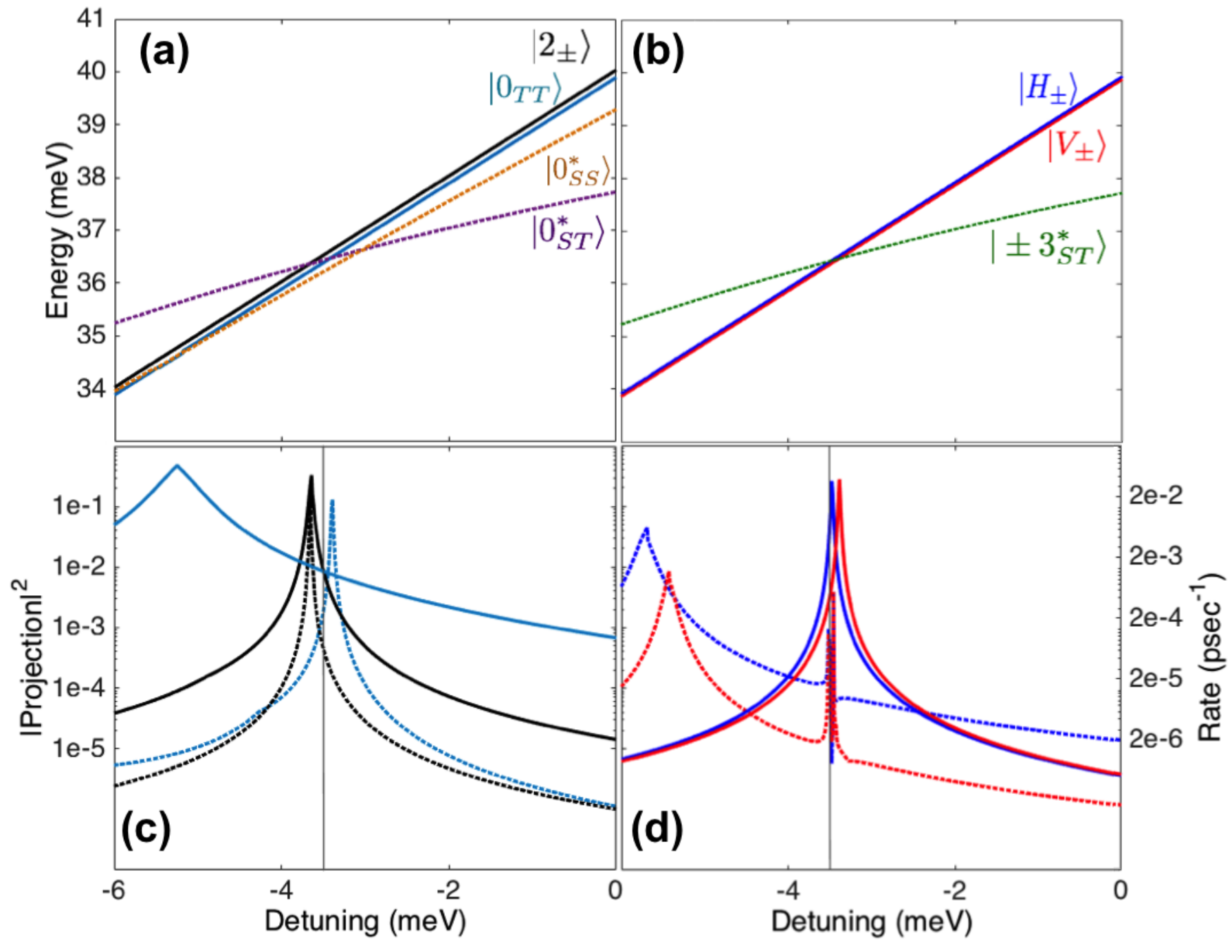} 
\caption{ (a) The calculated energies of the $|2_{\pm}\rangle$ (black), $|0_{TT}\rangle$ (teal), $|0^{*}_{ST}\rangle$ (purple dashed), $|0_{SS}^{*}\rangle$ (gold dashed) states  and  (b)  the calculated energies of the $|H_{\pm}\rangle$ (blue), $|V_{\pm}\rangle$ (red) and $|\pm3^{*}_{ST}\rangle$ (green dashed) states as a function of the detuning between the electron energy level separation $\Delta E_{1e2e}$ and the LO phonon energy $E_{LO}$. 
The corresponding calculated probabilities ($|$projection$|^2$) of having phonon-containing components in the wavefunction of the $|2_{\pm}\rangle$ state (black) and the $|0_{TT}\rangle$ state (teal) (c) and in the $|H_{\pm}\rangle$ states (blue), and the $|V_{\pm}\rangle$ states (red) (d). 
Solid lines represent probability resulting from a spin flip-flop relaxation pathway, and dashed lines represent probabilities resulting from a spin flip relaxation pathway. The scale to the right of (c) and (d) presents the calculated non-radiative relaxation rate. Note that the states are named by their leading terms at negative detuning, where the electron-LO phonon coupling is negligible.}
\label{fig:detuning}
\end{center}
\end{figure}

By diagonalizing the Hamiltonian in Eqn. \ref{eq:hamfull},
we obtain its eigenstates and eigenvalues, including the amount of mixing with the phonon-coupled states.  We can model this effect as a function of the detuning between the electronic level separation $\Delta E_{1e2e}$ and the LO phonon energy $E_{LO}$.  

Fig. \ref{fig:detuning} presents the energies of the various excited biexcitonic states as a function of the detuning between the $1e-2e$ electron energy level seperation, $\Delta E_{1e2e}$, and the LO phonon energy, $E_{LO}$. Fig. \ref{fig:detuning}(a) presents the states  $|2_{\pm}\rangle$, $|0_{TT}\rangle$, $|0^{*}_{ST}\rangle$ and $|0_{SS}^{*}\rangle$ that relax primarily to the 
biexcitonic states $|0_{SS}\rangle$ and $|0_{ST}\rangle$ and Fig. \ref{fig:detuning}(b) presents the states $|H_{\pm}\rangle$, $|V_{\pm}\rangle$ and $|\pm3^{*}_{ST}\rangle$ that relax primarily to the biexcitonic state $|\pm3_{ST}\rangle$. 

Fig. \ref{fig:detuning}(a), shows that, for a certain detuning, the $T^eT^h$ biexciton states, $|2_{\pm}\rangle$ and $|0_{TT}\rangle$ cross the $|0^{*}_{ST}\rangle$ and the $|0_{SS}^{*}\rangle$ states. When these states are nearly degenerate in energy, such that their energy separation is comparable to the exchange terms $\tilde{\Delta}_{1-}$, the spin flip-flop interaction becomes important. It significantly mixes between the phonon containing states and the zero-phonon state, thus providing an efficient non-radiative relaxation pathway to the $|0_{SS},1LO\rangle$ and $|0_{ST},1LO\rangle$ states. When the energy separation becomes comparable to the exchange terms ${\small{}\tilde{\Delta}_{(e,h)}}$ the spin-flip interactions also provide efficient non-radiative relaxation pathways. However, in the case of the spin-flip interactions, these relaxations are to the $|\pm 3_{ST},1LO\rangle$ states. 
Likewise, Fig. \ref{fig:detuning}(b) shows that, for almost the same detuning, the $|\pm3^{*}_{ST}\rangle$ state crosses the $|H_{\pm}\rangle$ states and $|V_{\pm}\rangle$ states. Here as well, when the energy separation between the states is comparable to the exchange terms, the spin-flip flop and the spin flip interactions again become important. In the case of the states in Fig. \ref{fig:detuning}(b), spin flip-flop allows for non-radiative relaxation paths to the $|\pm3_{ST},1LO\rangle$ states. The spin-flip interactions provide non-radiative relaxation paths to the $|0_{ST},1LO\rangle$ and $|0_{SS},1LO\rangle$ states. 

\begin{figure*}[tb]
\begin{center}
\includegraphics[width=\textwidth]{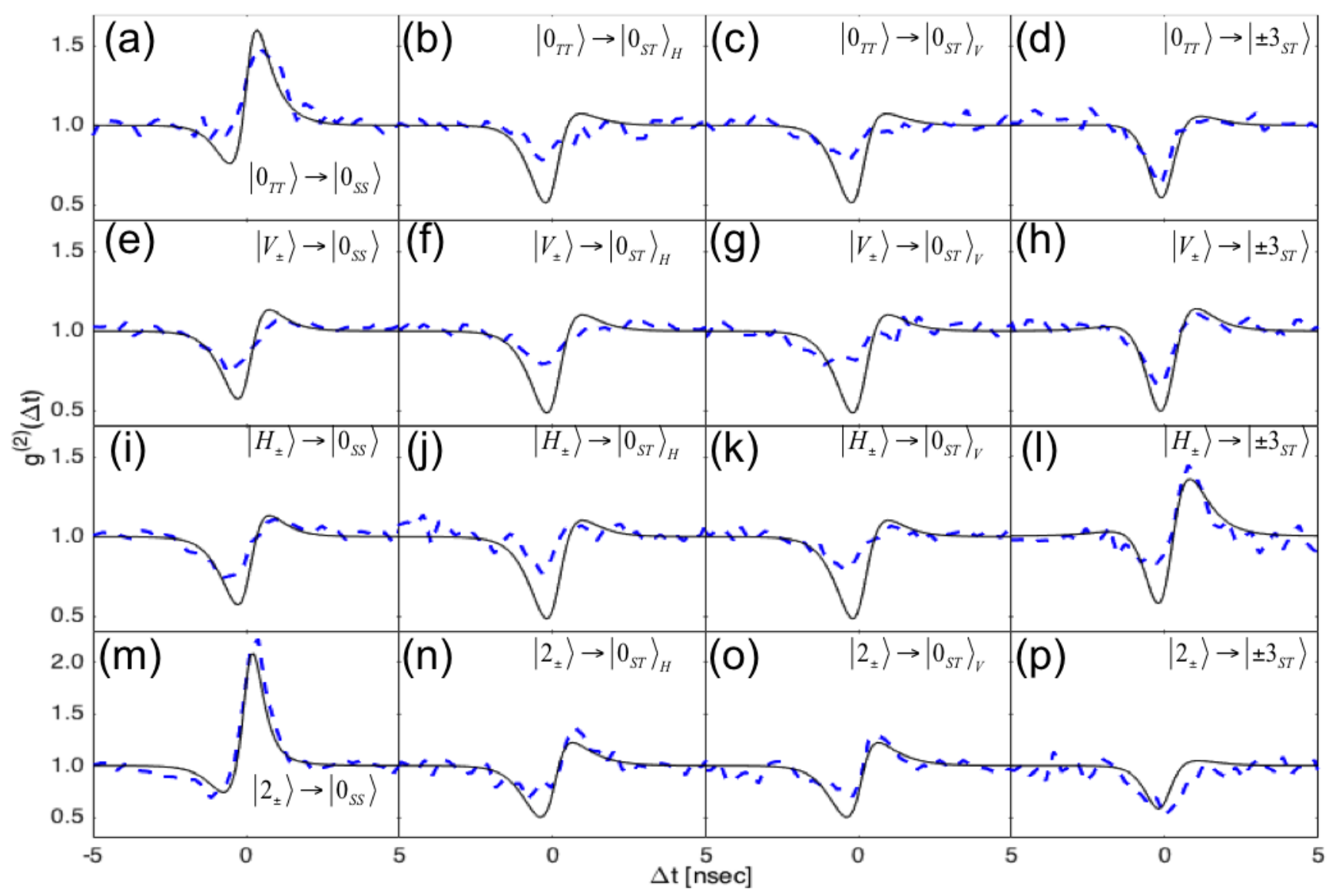}
\caption{Second-order intensity correlations between all triexcitonic emission lines (corresponding to different excited biexcitonic states) and all ground state biexciton lines. The studied non-radiative relaxations are as follows: (a-d)$|0_{TT}\rangle$ to $|0_{SS}\rangle$, $|0_{ST}\rangle$, and $|\pm3_{ST}\rangle$. (e-h) $|V_{\pm}\rangle$ to the same ground biexcitonic states. (i-l) $|H_{\pm}\rangle$ to the same ground biexcitonic states. (m-p) $|2_\pm\rangle$ to the same ground biexcitonic states.}
\label{fig:corrData}
\end{center}
\end{figure*}

Figs. \ref{fig:detuning}(c,d) show the phonon-containing components of the excited biexciton wavefunctions. Clear resonant enhancements are observed in the phonon containing part of the wavefunctions close to the crossing detuning. The solid lines indicate the contributions to the phonon-containing part of the wavefunction due to the spin flip-flop interaction. The dashed lines indicate the contributions due to the spin-flip interaction. The colors of the lines in Figs. \ref{fig:detuning}(c,d) match the colors of the states from Figs. \ref{fig:detuning}(a,b). 

The spin-flip exchange interaction (dark-bright exciton mixing) results in non-radiative relaxation pathways between the states in Fig. \ref{fig:detuning}(a) and the $|\pm 3_{ST},1LO\rangle$ radiating biexciton state, as well as relaxation pathways between the states in Fig. \ref{fig:detuning}(b) and the $|0_{ST},1LO\rangle$ and $|0_{SS},1LO\rangle$ radiating biexciton states. 

The calculated relaxation rates are determined by dividing the phonon containing probability by the lifetime of the LO phonon ($\sim$7 psec~\cite{li1999}).

\begin{table}[h]
{\footnotesize{}}%
\begin{tabular}{|c>{\centering}p{3cm}cc|}
 \hline
{\bf Parameter} & \multirow{1}{3cm}{{\bf Description}} & {\bf Value (meV)} & {\bf Ref.}\tabularnewline
\hline 
\hline 
$\Delta_{0}^{1,1}$ & Isotropic exchange between 1e and 1h & 0.271 & \cite{benny2014}\tabularnewline
\hline 
$\Delta_{0}^{1,2}$ & Isotropic exchange between 1e and 2h & 0.200 & \cite{benny2011prb}\tabularnewline
\hline 
$\Delta_{0}^{2,1}$ & Isotropic exchange between 2e and 1h & 0.200 & \cite{benny2011prb}\tabularnewline
\hline 
$\Delta_{0}^{2,2}$ & Isotropic exchange between 2e and 2h & 0.271 & \cite{benny2014}\tabularnewline
\hline 
$\Delta_{1}^{1,1}$ & Anisotropic exchange between 1e and 1h & -0.033 & \cite{benny2014}\tabularnewline
\hline 
$\Delta_{1}^{2,1}$ & Anisotropic exchange between 1e and 2h & 0.324 & \cite{benny2014}\tabularnewline
\hline 
$\Delta_{1}^{1,2}$ & Anisotropic exchange between 2e and 1h & 0.06 & \cite{benny2011prb}\tabularnewline
\hline 
$\Delta_{1}^{2,2}$ & Anisotropic exchange between 2e and 2h & 0.06 & \cite{benny2011prb}\tabularnewline
\hline 
$\Delta_{2}^{m,n}$ & Exchange between h at level $m$ and e at level $j$ & -0.0015 & \cite{benny2014}\tabularnewline
\hline
$\Delta^{(i,j)}_e$&Dark-bright mixing parameter for electrons&0.003&\tabularnewline
\hline
$\Delta^{(i,j)}_h$&Dark-bright mixing parameter for holes&0.003&\tabularnewline
\hline \hline
$E_{ic}$ & Energy of carrier $c$ in level $i$. & -15,-5,14,42\footnote{$E_{2h},E_{1h}, E_{1e}, and E_{2e}$ respectively} & \cite{benny2014}\tabularnewline \hline
$E_{g}$& Bandgap & 1297 &\cite{benny2014} \tabularnewline\hline
$E^{Coul}_{icjcjcic}$ & Direct Coulomb interaction between the carrier $c$ in the state $i$ and the carrier $c$ in state $j$. & 22.7, 17.0,26.3,19.7\footnote{$E^{Coul}_{1e1e1e1e}$, $E^{Coul}_{2e1e1e2e}$,$E^{Coul}_{1h1h1h1h}$,$E^{Coul}_{2h1h1h1h}$ respectively.} & \cite{benny2011prb, benny2014} \tabularnewline\hline
$E^{Coul}_{iejhjhie}$&Direct Coulomb interaction between e in state $i$ and h in state $j$&24.3,17.3,19.1,18.8\footnote{$E^{Coul}_{1e1h1h1e}$,$E^{Coul}_{2e1h1h2e}$, $E^{Coul}_{1e2h2h1e}$, and $E^{Coul}_{2e2h2h2e}$ respectively}&\cite{benny2011prb, benny2014}\tabularnewline\hline
$E^{exch}_{1e2e1e2e}$&e-e exchange interaction&3.7 &\cite{benny2011prb}\tabularnewline\hline
$E^{exch}_{1h2h1h2h}$&h-h exchange interaction&6.6&\cite{benny2011prb}\tabularnewline \hline
$C_{F}$ & Fr{\"o}hlich coupling constant  & 6.4 & \cite{benny2014}\tabularnewline
\hline 
$E_{LO}$ & LO phonon energy & 32  & \cite{benny2014}\tabularnewline 
\hline
$E_{1e-2e}$& Electron $1e-2e$ energy splitting & 28.5  & \cite{benny2014}\tabularnewline
\hline
\end{tabular}\protect\caption{List of parameters used in calculations.\label{tab:parameters}}
\end{table}

Figs. \ref{fig:detuning}(b,d) show that the combination of the electron-LO phonon interaction and the e-h exchange interaction result in substantial enhancement of the non-radiative relaxation rates to various ground biexcitonic states. These rates crucially depend on the precise value of the detuning between the electron energy level separation and the LO phonon energy. The vertical line in Figs. \ref{fig:detuning}(b,d), corresponding to a value of -3.5 meV, represents a detuning where the radiative relaxation rates match those observed in the specific QD studied in this work. This detuning value corresponds well with the value of approximately -4 meV reported in Ref. \cite{benny2014}. 

\section{Experimental Results and Discussion}
\label{sec:exp}

From the above analysis, we see that the LO phonon primarily allows for non-radiative relaxation from $|H_{\pm}\rangle$ to $|\pm3_{ST}\rangle$, from $|2_{\pm}\rangle$ to $|0_{SS}\rangle$ and $|0_{ST}\rangle$, and from $|0\rangle$ to $|0_{SS}\rangle$. We have identified the spectral lines corresponding to recombination from the triexcitonic states (Fig. \ref{fig:theory}(d)) and the spectral lines corresponding to recombination from the $|0_{SS}\rangle$, $|0_{ST}\rangle$ and $|\pm3_{ST}\rangle$ states. Fig. \ref{fig:PL} identifies these lines in the photoluminescence emission of a single QD under nonresonant excitation with a 455 nm cw laser.  The spectral lines have been
identified by their excitation intensity dependence, PL excitation
spectra \cite{benny2011prb}, polarization dependence \cite{poem2007},
and for the case of the triexcitonic emission lines, by second-order
and third-order intensity correlation measurements with both the neutral
biexciton and the neutral exciton.\cite{schmidgall2014}.

\begin{table}
\begin{tabular}{|cc|}
\hline
Parameter & Value \\ \hline \hline
Detuning & -3.5 meV \\ \hline
Radiative lifetime & 400 ps \\ \hline
$|H_{\pm}\rangle \rightarrow |\pm3_{ST}\rangle$ & 50 ps\\ \hline
$|V_{\pm}\rangle \rightarrow |\pm3_{ST}\rangle$ & 3000 ps\\ \hline
$|2_{\pm}\rangle \rightarrow |0_{SS}\rangle$ & 400 ps \\ \hline
$|2_{\pm}\rangle \rightarrow |0_{ST}\rangle$ & 1000 ps \\ \hline
$|0_{\pm}\rangle \rightarrow |0_{SS}\rangle$ & 1200 ps \\ \hline
$|H_{\pm}\rangle, |V_{\pm}\rangle \rightarrow |0_{SS}\rangle$ & 5000 ps \\ \hline
\end{tabular}
\caption{Parameters used to fit the measured correlation function, determined by the phonon projection at the selected detuning.}
\label{tbl:rate_params}
\end{table}

To study the non-radiative relaxations of the excited biexcitonic states, we perform polarization sensitive two-photon intensity correlation measurements, where the first photon results from recombination of a triexcitonic state and the second photon results from recombination of a ground or a metastable biexciton state. 
The second-order intensity correlation
function is given by 
\begin{equation}
g_{1,2}^{(2)}(\tau)=\langle I_{1}(t)I_{2}(t+\tau)\rangle/\left(\langle I_{1}(t)\rangle\langle I_{2}(t)\rangle\right)
\end{equation}
 where $I_{i}(t)$ is the intensity of light at time $t$ on the $i$th
detector, $\tau$ is the time between the detection of a photon in
detector 1 and detection of a photon in detector 2, and $\langle\rangle$
means temporal average. Following the detection of the second photon
in a cascade, no detection of emission from the first photon is possible.
Therfore, ``antibunching'' {[}$g^{(2)}(\tau)<1]$ is observed. However,
following the detection of the first photon, the probability of detecting
the second photon is higher than the steady state probability \cite{dekel2000,regelman2001,poem2012}
and bunching {[}$g^{(2)}(\tau)>1]$ is observed. The observation of
bunching in second-order intensity correlation measurements between
the emission resulting from recombination of the triexciton and the emission resulting from the various biexciton states indicates
that the non-radiative relaxation rate from the e-triplet-h-triplet
excited biexciton states to the various radiating biexciton states is comparable to the radiative rates.

The second order intensity correlation function was measured for each triexciton
emission line and each of the biexcitonic emission lines, for a total
of sixteen measurements. The QD was excited with a non-resonant 445nm cw diode laser, and the excitation power was chosen such that the PL intensity of the $XX^{0}$ and $X^0$ emission lines were equal. These measurements are presented in Fig. \ref{fig:corrData}, and labeled with the relevant non-radiative relxation. The second-order intensity functions have been fit using a rate equation model. 

Generally, modeling PL and second-order intensity correlation functions requires calculation of all elements of the density matrix governing the quantum system as a function of time. However, since the rates of environmentally-induced dephasing are faster than the time scale of the dynamics between the QD states, the off-diagonal elements of the density matrix can be neglected, and we can simulate the QD population dynamics using a set of coupled classical rate equations, $\frac{d\vec{n}(t)}{dt} = \overleftrightarrow{R} \vec{n}(t)$, where $\vec{n}(t)$ is a vector of state probabilities and $\overleftrightarrow{R}$ is a matrix of time-independent transition rates between the states, and an excitation intensity dependent generation rate.\cite{regelman2001} The transition rates taken into account in these rate equations include the non-radiative processes outlined above, with the rates as determined from the above model, and the radiative lifetime of the states (400 ps) \cite{ICPS}. The time dependent correlation functions are calculated by solving the coupled rate equations with initial conditions set by the detected first photon. The probability to detect a particular second 
photon as a function of time, is then found from the time dependent probability to populate the state from which the photon is emitted.  The fit to the measurements is achieved by convoluting the calculated correlation function with the temporal response of the photodetectors in the experimental 
setup. 
The parameters used in the rate equation model, including the rates extracted from the phonon coupling, are presented in Table \ref{tbl:rate_params}. 

\begin{figure}[tb]
\begin{center}
\includegraphics[width = \columnwidth]{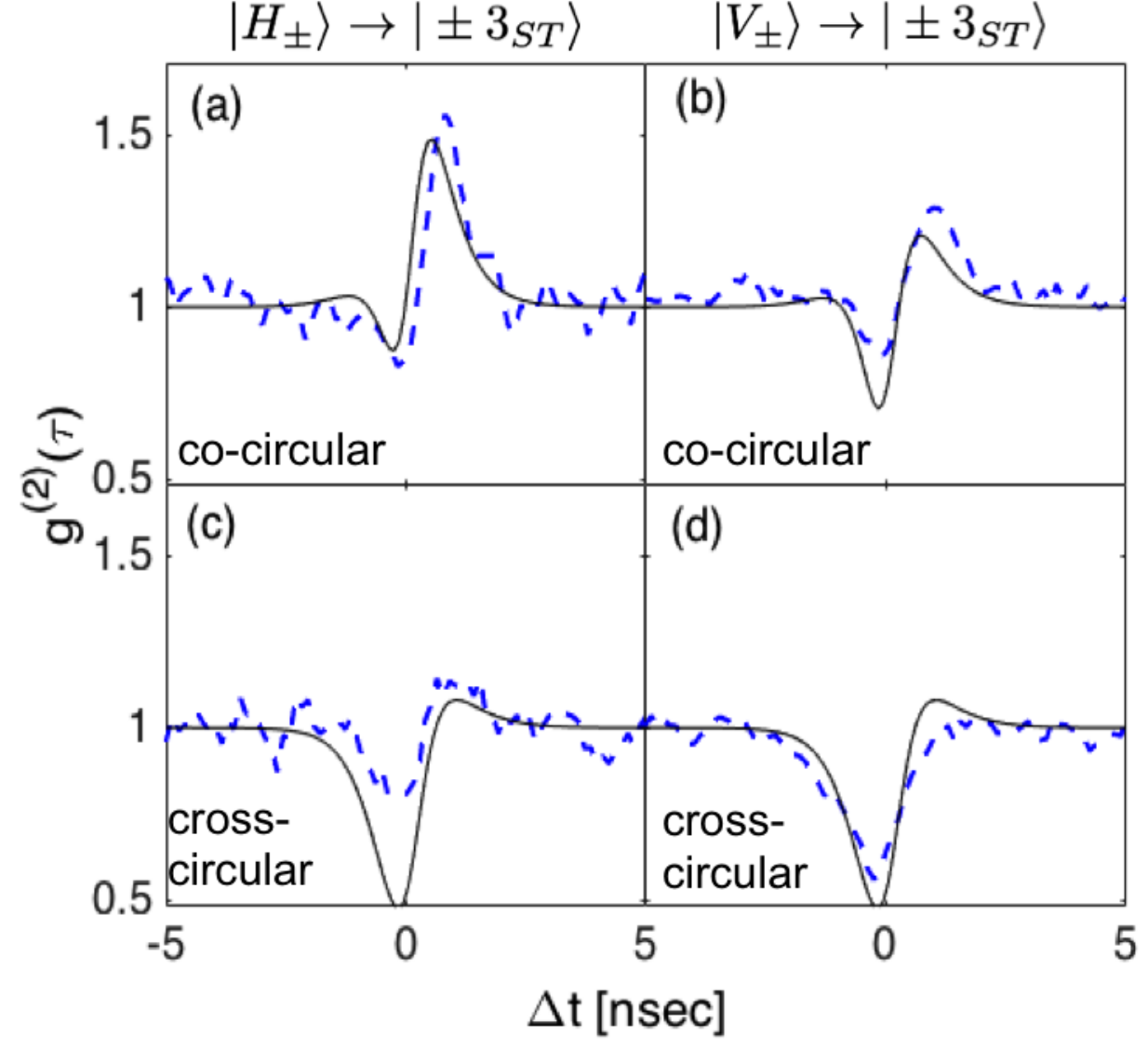}
\caption{Second-order intensity correlations between the triexcitonic emission lines initiating in the $|H_{\pm}\rangle$ (a,c) [$|V_{\pm}\rangle$ (b,d)] state and the $XX^0_{T\pm3}$ biexcitonic emission line, indicating non-radiative relaxation to the $|\pm3_{ST},1LO\rangle$ state. (a,b) The two photons are co-circularly polarized. (c,d) The two photons are cross-circularly polarized.}
\label{fig:circ}
\end{center}
\end{figure}

These second-order intensity correlation measurements verify the predictions of the model. First, the $|0_{TT}\rangle$ state relaxes primarily to the $|0_{SS}\rangle$ state. Secondly, the $|2_{\pm}\rangle$ states relax both to the $|0_{SS}\rangle$ and $|0_{ST}\rangle$ states. Finally, we see that the $|H_{\pm}\rangle$ and $|V_{\pm}\rangle$ states relax primarily to the $|\pm3_{ST}\rangle$ state, and that the relaxation of the $|H_{\pm}\rangle$ state is substantially faster than that of the $|V_{\pm}\rangle$ state. This faster relaxation can be observed in the stronger maximal bunching signal ($g^{(2)}_{max} \approx 1.5$) observed in the $|H_{\pm}\rangle \rightarrow |\pm3_{ST}\rangle$ cascade (Fig. \ref{fig:corrData}(l)) when compared to the maximal bunching signal ($g^{(2)}_{max} \approx 1.2$) observed in the $|V_{\pm}\rangle \rightarrow |\pm3_{ST}\rangle$ cascade (Fig. \ref{fig:corrData}(h)). In all cases, the model-obtained rates result in a reasonable fit of the rate equation model to the bunching in the experimental data.  In several cases, the model overpredicts the amount of antibunching. The depth of the modeled antibunching is determined by two values: the response time of the detector and the carrier generation rate, where longer response times and/or higher rates  lead to shallower antibunching. The modeled generation rate was selected such that the modeled intensity of the $XX^{0}$ and $X^{0}$ emission lines was equal, corresponding to the laser power used in the experiment. 

The data also indicate that there is a slow non-radiative relaxation pathway from $|H_{\pm}\rangle$ and $|V_{\pm}\rangle$ to $|0_{SS}\rangle$, which results from the mixing between the BE and the DE in the model, as indicated in Fig. \ref{fig:detuning}(c). In this case, a dark-bright mixing term of $3 \mu$eV for both the electrons ($\Delta_e$) and the holes ($\Delta_h$) was used \cite{slava2016}. This results in a non-radiative relaxation rate of approximately 5 nsec, a value which matches the observed bunching in Fig. \ref{fig:corrData}(e,i). The DE optical depletion experiments reported in Ref.\cite{schmidgall2015} rely on this relaxation channel, and they suggest a similar, few-nsec relaxation rate.  Thus, the dark-bright mixing values in the model result in predicted rates that correspond well to both the observed bunching in Figs. \ref{fig:corrData}(e,i) and previous experimental data. 

As an example for the strength of the experimental technique in determining the non radiative decay mechanisms and their rates we present in  Fig. \ref{fig:circ} circular polarization sensitive intensity correlation measurements. In \ref{fig:circ} we measure the correlations between photons emitted from the linearly polarized ``dark-like'' triexciton emission lines associated with the $|H_{\pm}\rangle$ states (line (ii) in the inset to Fig. 2) and the $|V_{\pm}\rangle$ states (line (iii) in the inset to Fig.2) and photons emitted from the unpolarized biexciton emission line associated with the $|\pm 3_{ST},1LO\rangle$ states ($XX^{0}_{T\pm3}$ in Fig. 2). In Fig. \ref{fig:circ} (a) and (b) the $|H_{\pm}\rangle$ and $|V_{\pm}\rangle$ triexciton photons respectively are co-circularly polarized with the biexciton photons and in Fig. \ref{fig:circ} (c) and (d) they are cross-circularly polarized. 

By inspecting Fig. \ref{fig:theory}(a) one sees that detecting a $\sigma^{+}$ [$\sigma^{-}$] circularly polarized photon from either of the two ``dark''-like triexciton states indicates that the excited biexciton state contains $|+1\rangle$ and $|-3\rangle$ [$|-1\rangle$ and $|+3\rangle$] components. Similarly, by inspecting Fig. \ref{fig:theory}(c), one sees that detecting a $\sigma^{+}$ [$\sigma^{-}$] circularly polarized photon from the $XX^{0}_{T\pm3}$ biexciton 
indicates that the emitting ground biexciton state is the  $|+3_{ST}\rangle$ [$|-3_{ST}\rangle$]. 
The non radiative process that connects between the initial, excited biexciton state and final, ground biexciton state must therefore proceeds via the e-h spin flip-flop process which accompanies the phonon-assisted electronic relaxation.    
During this process the $|+1\rangle$ [$|-1\rangle$] component of the state transforms into $|+3_{ST}\rangle$ [$|-3_{ST}\rangle$], state resulting eventually in the emission of a second $\sigma^{+}$ [$\sigma^{-}$] co-polarized photon. This is clearly seen in the bunching signal observed in the co-circular polarization measurements of Fig. \ref{fig:circ} (a) and (b) and its absence in the cross-circular polarization measurements of Fig. \ref{fig:circ} (c) and (d). All the measurements agree well with the rates obtained from the model (Table \ref{tbl:rate_params}). 

\section{Conclusions}

In summary, we measure non-radiative relaxation processes in two-photon radiative cascades initiating from the quantum-dot confined triexciton. These radiative cascades include non-radiative relaxation pathways which do not conserve spin, resulting from the electron-hole exchange interaction in the presence of subband electronic level separation near resonance with LO phonon. We demonstrate quantitative agreement between 16 different two photon intensity correlation measurements and a model which includes the electron-hole exchange and the electron-phonon interaction. In particular, we show that even small electron-hole exchange terms may be significantly enhanced under these resonant conditions. The quantitative and qualitative
understanding of this phenomenon may enable the engineering of deterministic
spin flip and flip-flop processes in semiconductor nanostructures. 

\begin{acknowledgments}
The support of the Israeli Science Foundation (ISF),
the Technion's RBNI and the Israeli Nanotechnology Focal Technology Area on "Nanophotonics for Detection" are gratefully acknowledged.
\end{acknowledgments}

\bibliographystyle{unsrtnat}
\bibliography{master_bibtex_db}

\end{document}